\documentclass[
	twocolumn,
	amsmath,
	amssymb,
	nofootinbib,
	]{revtex4-2}
\usepackage{graphicx}
\usepackage{xcolor}
\usepackage{amsmath}
\usepackage{amssymb}
\usepackage{booktabs}
\usepackage{titlesec}

\begin{document}

\title{Counting and Sequential Information Processing in Mechanical Metamaterials}

\author{Lennard J. Kwakernaak}
\affiliation{Huygens-Kamerlingh Onnes Laboratory, Universiteit Leiden, PO Box 9504, 2300 RA Leiden, the Netherlands}
\affiliation{AMOLF, Science Park 104, 1098 XG Amsterdam, the Netherlands}
\author{Martin van Hecke}
\affiliation{Huygens-Kamerlingh Onnes Laboratory, Universiteit Leiden, PO Box 9504, 2300 RA Leiden, the Netherlands}
\affiliation{AMOLF, Science Park 104, 1098 XG Amsterdam, the Netherlands}
\
\date{\today}

\begin{abstract}
Materials with an irreversible response to cyclic driving
exhibit an evolving internal state which, in principle, encodes information on the
driving history. Here we realize {irreversible metamaterials that count} mechanical driving cycles and store {the result} into easily interpretable internal states. We extend these designs to aperiodic metamaterials which are sensitive to the order of different driving magnitudes, and realize 'lock and key' metamaterials that only reach a specific state
for a given target driving sequence. Our strategy is robust, scalable and extendable, and opens new routes towards smart sensing, soft robotics and mechanical information processing.
\end{abstract}

\maketitle

Counting a series of signals is an elementary process that can be materialized in simple electronic or neural networks \cite{Amit1988}.  Even the Venus flytrap can count, as it only snaps shut when touched twice, despite not having a brain \cite{Bohm2016}.
While the ability to count
is not commonly associated with materials, certain complex materials, from crumpled sheets to amorphous media, can exhibit memory effects where the state
depends on the driving history \cite{mungan2019networks,keim2019memory}.
Under cyclic driving, their response then may feature subharmonic behavior {\cite{multi1,multi2,multi3,multi4,multi5,multi6,multi7,keim2021multiperiodic}} or, as was recently shown, a transient where the system only settles in a periodic response after $\tau>0$ driving cycles \cite{Lind2021,hbense,yoav}.
The latter response thus counts the number of driving cycles in principle,
but in practice, the link between this number and the internal state is highly convoluted.
Materials that would feature controlled counting could simplify the design of soft robotics and intelligent sensors, and more widely, open a route towards sequential information processing.
{However, we have no rational strategies to control the link between state
and count or to realize
in-material counting.}

\begin{figure}[t]
	\centering
	\includegraphics{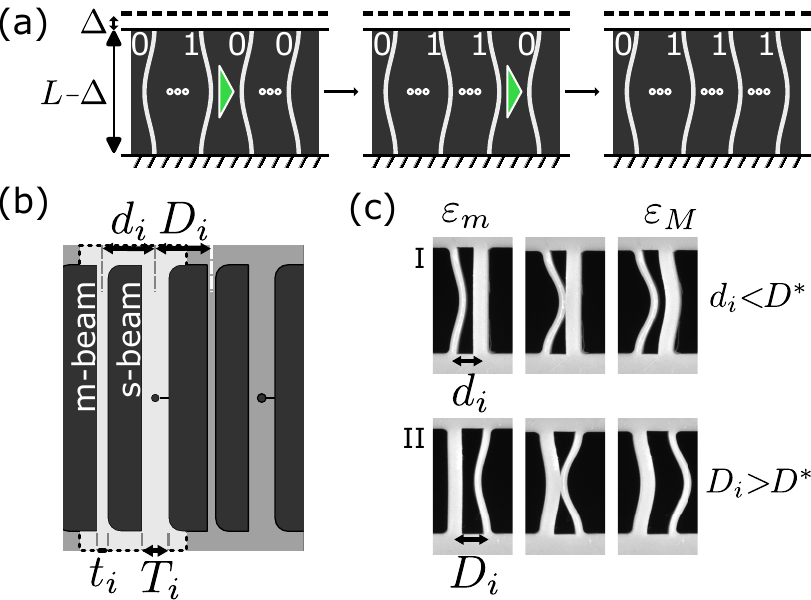}
\caption{(a) Schematic representation of the evolution of a 'beam counter' metamaterial
with $n=4$ unit cells under two compression cycles.
Each unit cell contains a buckled m-beam (memory beam) which encodes a single bit.
When the strain $\varepsilon:=\Delta/L$ is cycled between $\varepsilon_m$ and $\varepsilon_M$, the m-beams interact (grey symbols), so that
the '1' state is copied to the right (triangle), leading to the step wise advancing of the '1' state to the right.
(b) Geometry of a unit cell $i$ (highlighted), containing a slender m-beam and a thicker, asymmetric s-beam ({slitted beam}) --- lengths are non-dimensionalized by setting the beams rest-lengths to 1.
(c) Evolution of {beam pairs} under increased compression from $\varepsilon_m$ (left) to $\varepsilon_M$ (right): when the spacing $d_i$ is smaller than a critical distance $D^*$, {the} buckled state of the slender beam is copied to the thick beam
(top); when $D_i>D^*${,}
the buckled state of the thick beam is copied to the slender beam (bottom).
	}
	\label{fig:1}
\end{figure}

Here we introduce a general platform for metamaterials  \cite{bertoldi2017flexible}
that count mechanical compression cycles. Our metamaterials consist of unit cells that each feature a memory-beam (m-beam) that is either buckled left or right, which we represent with a binary value $s_i=0$ or 1 \cite{pedro} (Fig.~\ref{fig:1}a). The unit cells are designed to interact with their neighbors such that under cyclic compression
any unit cell in the '1' state copies this state to its right neighbor (Fig.~\ref{fig:1}b-c). This  leads to a mechanically clocked wave where the '1' state advances rightward, one unit cell per compression cycle. Hence, the collective state, $S:=\{s_1,s_2,\dots\}$, evolves like in a cellular automaton \cite{automata}, with repeated cyclical compression yielding { simple predictable pathways.}

We combine such beam counters to realize metamaterials which exhibit more complex forms of sequential information processing than counting, including the
detection of compression cycles of multiple amplitudes, as well as their sequential order. Together, these establish a general platform for realizing targeted multi-step pathways in metamaterials and open a route towards sequential information processing {\em in materia} \cite{rise_of_intelligent_matter,nature_mech_logic_gates,mechanical_computing_review}.

\begin{figure}[t]
	\centering
	\includegraphics{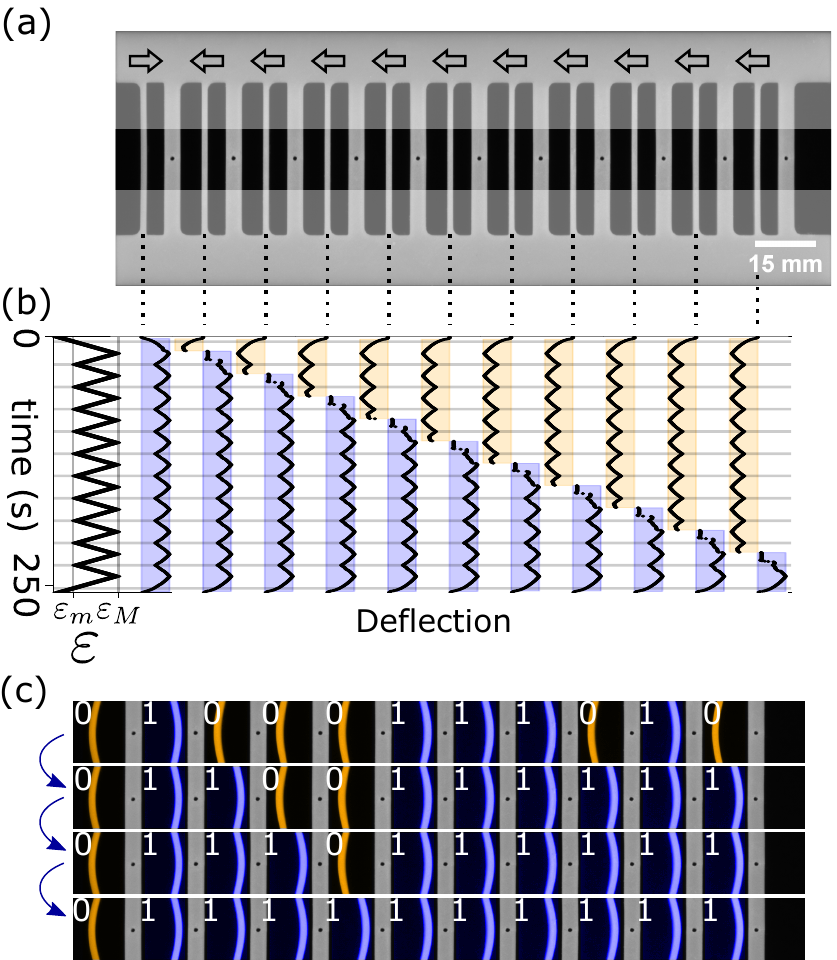}
	\caption{
(a) Beam counting metamaterial with 11 units, with center region highlighted; see Movie 1 in \cite{SI} ($t_i=0.040$, $T_i=0.10$, $d_i=0.13$ and $D_i=0.15$).
Arrows indicate weak symmetry breaking of the m-beams that makes the system reach state $\{10000000000\}$ when $\varepsilon$ is increased from zero.
(b) Space-time plot, tracing the center positions of each m-beam as a function of time under cyclic compression $\varepsilon_m \nearrow \varepsilon_M \searrow \varepsilon_m$
($\varepsilon_m=0.026$, $ \varepsilon_M=0.099$). Beams in state 0 (1) are highlighted in yellow (blue).
(c) Evolution of the beam counter prepared in the initial state  $\{01000111010\}$ (central parts shown only; beam state colored as above).}
	\label{fig:2}
\end{figure}

\section{Unit cell and cyclic driving}
We aim to realize metamaterials where state '1' spreads to the right when the compressive strain $\varepsilon$ is cycled between $\varepsilon_m$ and $ \varepsilon_M$ (Fig.~\ref{fig:1}).
We note that in contrast to recent metamaterials which exhibit sequential shape changes under monotonic driving \cite{overvelde2015amplifying,coulais2018,murugan2018,
kresling_2021,melancon}, we require a sequential response under cyclic driving.
This necessitates unit cells that memorize their previous state, interact with their neighbors, and break left-right symmetry. We satisfy these requirements with unit cells $i$ containing two beams (Fig.~\ref{fig:1}b). The slender m-beams encode states $s_i=0$ or 1 in {their} left or right buckled configurations. We choose  $\varepsilon_m$ larger than their buckling strain so they retain their state.
The thicker and non-trivially shaped s-beams facilitate interactions between the m-beams, and buckle at a strain larger than $\varepsilon_m$ but smaller than $\varepsilon_M$.

The detailed design involves a careful choice of the symmetry breaking beam shapes and their spacings.
First, weakly symmetry breaking rounded corners at the ends of the m-beams controls their
buckling into a desired initial configuration $S=\{100\dots\}$ --- this does not appreciably modify the evolution of the sample during compression cycles, yet allows resetting the beam counter by momentarily cycling the strain towards zero.
Second, the s-beams feature similarly rounded corners that makes them buckle left,
and a slit which extends their reach when
they snap  to the right and the slit opens up \cite{bernatpaper}.
As we show below,  these symmetry breaking enhancements are crucial for their role in right-copying the '1'-state of the m-beams.
Third, we use the beam spacings $d$ and $D$ to control the interactions between s- and m-beams. We found that when two buckled beams of unequal thickness are brought in contact, upon further compression they either both snap left or snap right --- the direction depends on whether their distance is smaller or larger than a critical distance $D^*$. We choose $d_i<D^*$ and $D_i>D*$ so that contact interactions between neighboring m- and s-beams favor rightward snapping of the beams (Fig.~\ref{fig:1}c).

\begin{figure}
	\includegraphics{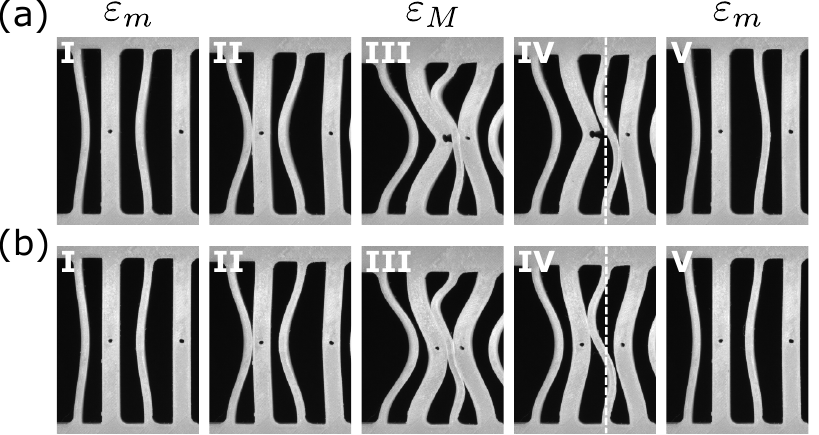}
	\caption{
	Comparison of the evolution of two unit cells during a compression cycle.
(a) Original design. (b) Design without slits, which does not copy the '1' state.
	Frames (aIV) and (bIV) are not at the same strain, but compare the states where the
second m-beam just loses contact with one of its neighboring s-beams;
note that the m-beam is then, respectively, to the right (a) and left (b) of the neutral line (dashed).
		}\label{fig:3}
\end{figure}

\section{Counting and controllable transients}
We combine our unit cells to construct a 'beam counter'
with $n=11$ unit cells, using standard 3D printing and molding techniques
(see Supplemental Material \cite{SI}; Fig.~\ref{fig:2}). We cycle the compression in a custom built setup that allows accurate parallel compression of wide samples, and track the center locations of the middle of the m-beams (Fig.~\ref{fig:2}b).
Ramping up the strain from zero to $\varepsilon_m$, the system reaches the initial state $\{10000000000\}$ (Fig.~\ref{fig:2}b). Repeated compression cycles
show the step-by-step copying of the '1'state of the m-beams to the right, which involves rightward snapping of the appropriate m-beam just after $\varepsilon$ has peaked (Fig.~\ref{fig:2}b).
Hence, the state evolves as
$\{100\dots0\}\to\{110\dots0\}\to\dots\to\{111\dots1\}$ (Fig.~\ref{fig:2}b)
(see Movie 1 in \cite{SI}).  We characterize such 'domain wall' states consisting of a string of 1's followed by 0's by the number of 0's, $\sigma$.
The evolution of our beam counter under cyclic compression can thus be seen as
as counting down from $\sigma=10$ to $\sigma=0$.
Our design is robust, can be scaled down, and can be operated in a hand-held device (see Movie 2 in \cite{SI}).

\begin{figure*}[t]
	\includegraphics{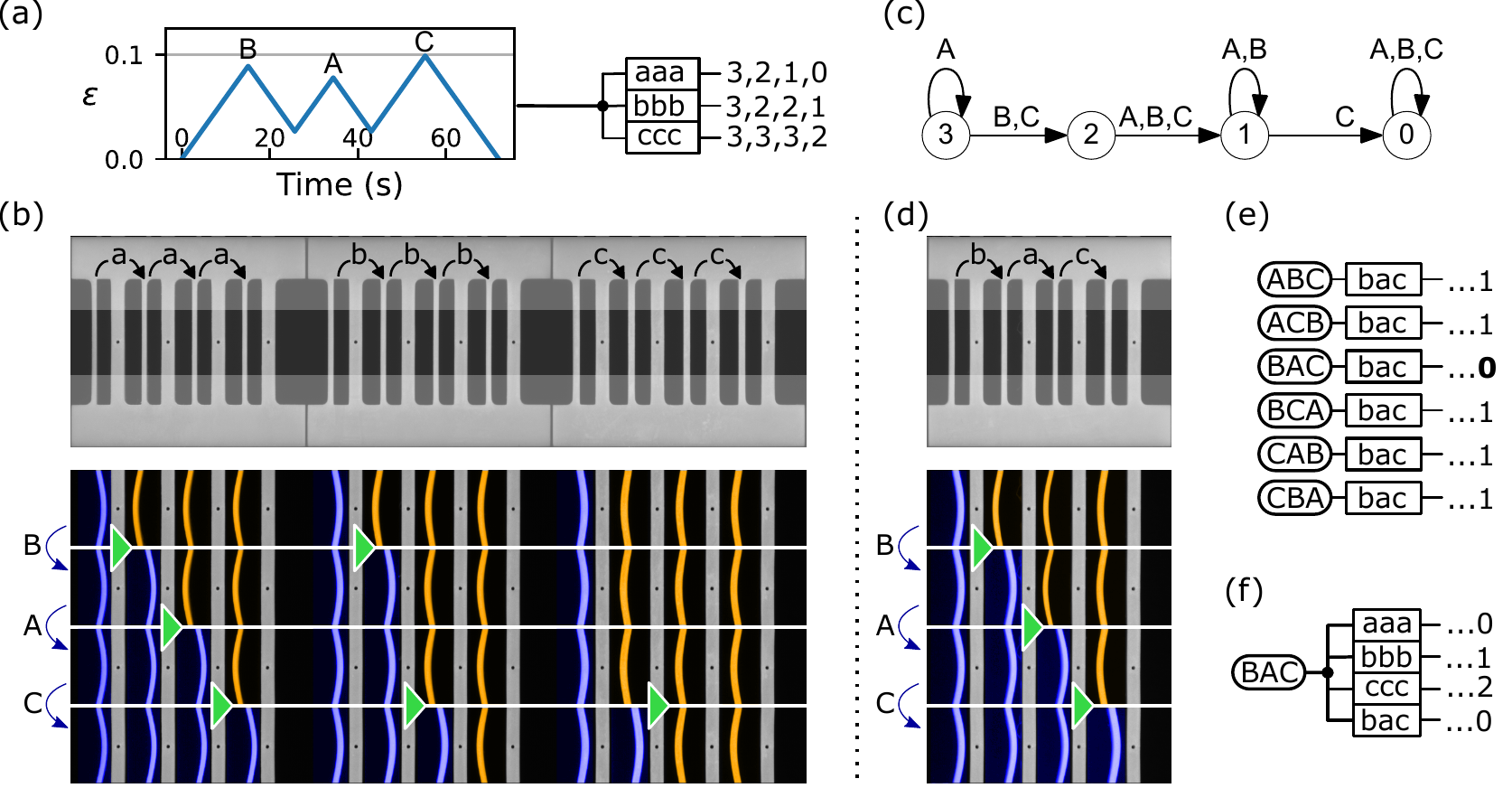}
	\caption{Response of aggregate metamaterials containing
multiple homogeneous and heterogeneous beam counters.
		(a) Complex driving cycles can be encoded as an input string (here $BAC$) and produce different sequential response in counters $aaa$, $bbb$ and $ccc$ {(boxes represent individual counters and numbers represent the sequence of states $\sigma_i$)}.
		(b) Experimental realization of the $aaa|bbb|ccc$ {aggregate} metamaterial, and snapshots of its pathway in response to  input driving $BAC$ (triangles highlight the right copying of a 1-bit). Note that $t_i$, $d_i$ and $D_i$ subtly vary along the counters while $T_i$ remains constant (see Supplemental Material \cite{SI}).
		(c) Transition graph of the heterogeneous counter $bac$.
		(d) Experimental realization and evolution of
$bac$ counter during input string $BAC$ (panel (b) and (d) are snapshots of the same experiment).
		(e) The output of the $bac$ counter to all input sequences which are permutations of $ABC$ (rounded boxes).
(f) The four counters $aaa$, $bbb$, $ccc$ and $bac$ operated in parallel reach the unique state $\{0120\}$ only for input sequence
		$BAC$. The homogeneous counters {store} the {count} of each signal, while the heterogeneous counter is sensitive to the specific permutation.
	}\label{fig:4}
\end{figure*}

The evolution from the natural initial state $\{100\dots\}$ only features a limited set of states, which do not contain substrings like $010$ or $001$. To demonstrate that our metamaterial correctly copies 1-bits to the right,
we use manual manipulation to program the metamaterial in the initial state $\{01000111010\}$ --- this state contains all possible three-bit substrings.
Its evolution shows that our metamaterial faithfully executes our target evolution (Fig.~\ref{fig:2}c; see Movie 3 in \cite{SI}).
Moreover, we note that this initial state evolves to the absorbing state $\{11\dots\}$ after only $\tau=3$ cycles, as the largest numbers of 0's to the right of a 1 is equal to three. Here, the transient $\tau$ is not a material property but a simple function of the state \cite{Lind2021,hbense}.

A detailed inspection of the evolution of adjacent unit cells during their
evolution illustrates that bit-evolution takes place in a
two phases (Fig.~\ref{fig:3}). First, when $ \varepsilon$ is increased beyond a unit-cell dependent critical strain $ \varepsilon^\dagger$,
the '1' state of $m_i$ is copied to $s_i$ (Fig.~\ref{fig:3}aI-aIII).
During this first phase, the left s-beam snaps open to the right, and the m-beam becomes sandwiched between two s-beams (Fig.~\ref{fig:3}aIII). In the second phase,
$\varepsilon$ is lowered to $\varepsilon_m$, and the sandwiched m-beam snaps right, after which all beams relax to their new configuration (Fig.~\ref{fig:3}aIII-aV).
To illustrate how the slits facilitate the copying of the right-buckled state, we compare the sandwiched states for s-beams with and without slits ((Fig.~\ref{fig:3}); see Movie 4 and Supplemental Information \cite{SI}). Without slits, the sandwiched m-beam is pushed left and first loses contact with the right beam; with slits, the m-beam is pushed right, first loses contact with the left beam, and eventually moves right  (Fig.~\ref{fig:3}a-b).
We stress that although the slits are essential in the current design,
we also realized beam counting in an alternative design that does not feature
slitted beams (see Movie 5 and Supplemental Information \cite{SI}).

\section{Sequential Processing}
To demonstrate process information beyond simple counting, we combine multiple beam counters into aggregate metamaterials (Fig.~\ref{fig:4}).
Our first goal is to realize metamaterials which discriminate and count driving cycles of different peak compressions $\varepsilon_M$.
Specifically,
we combine three $n=4$ beam counters labeled
$aaa, bbb$, and $ccc$
which have respective critical thresholds
$(\varepsilon_a^\dagger,\varepsilon_b^\dagger,\varepsilon_c^\dagger) \approx (0.073(4), 0.085(3), 0.092(2))$,
which are all controlled by the same global strain $\varepsilon$ (Fig.~\ref{fig:4}a-b).
We label the resulting metamaterial as  $aaa|bbb|ccc$, and
characterize its  state by the number of '0' beams
in each counter, $\{\sigma_i\}$.
We define driving cycles of different magnitude, $A,B,C$, as compression sweeps $\varepsilon_m \nearrow \varepsilon_M^{A,B,C}\searrow \varepsilon_m$, with
$\left(\varepsilon^A_M, \varepsilon^B_M, \varepsilon^C_M \right) \approx (0.078, 0.089, 0.099)$, such that $\varepsilon_a^\dagger<\varepsilon_M^A<\varepsilon_b^\dagger< \varepsilon_M^B<\varepsilon_c^\dagger<\varepsilon_M^C$.
Starting out in the initial state $\{\sigma_i\}=\{3,3,3\}$, a single driving cycle ($A$, $B$ or $C$) then advances
one, two or all three counters, yielding three distinct states $\{2,3,3\}$, $\{2,2,3\}$, or $\{2,2,2\}$ respectively. Hence, from the state we can uniquely infer the applied driving cycle.

Crucially, longer driving sequences are also encoded in the internal state
 We denote  sequential driving cycles as, e.g., $BAC$, for which
$\{\sigma_i\}$ evolves as $\{3,3,3\}\xrightarrow{B}\{2,2,3\}\xrightarrow{A}\{1,2,3\}\xrightarrow{C}\{0,1,2\}$ (Fig.~\ref{fig:4}b). These states all encode specific information, e.g., state $\{1,2,3\}$ encodes one $A$ and one $B$ pulse, whereas $\{0,1,2\}$ encodes a memory of one $B$, one $C$ and an arbitrary number of $A$ pulses. We note that while
the capacity of our metamaterial is limited by
one or more counters reaching zero, it can be enlarged by increasing
the length $n$ of the counters. Furthermore, we note that our metamaterial precisely materializes the Park Bench model that has been introduced as a toy model to understand Multiple Transient Memories \cite{paulsen2014multiple,paulsen2019minimal}. Regardless, our strategy combining multiple beam counters allows
to distinguish and count different signals.

So far, our metamaterials are insensitive to the order of input signals, which limits their functionality to counting. However, combining unit cells with different thresholds in a single 'strip' realizes heterogeneous metamaterials whose response is sequence dependent and, e.g., discriminates driving cycles $ABC$ from $BAC$.
We realize the heterogeneous metamaterial $bac$. (Fig.~\ref{fig:4}c-e).
Starting from state $\sigma=3$, we can use the same logic as before to infer
its evolution and we subsequently collect all possible pathways in a transition graph (Fig.~\ref{fig:4}c). In particular we find that input $BAC$ yields $\sigma=0$
while all other three-character permutations of $A$, $B$ and $C$ yield $\sigma=1$ (Fig.~\ref{fig:4}d-e). This illustrates that the response of  heterogenous counters is sequence dependent.

Finally, by combining heterogeneous and homogeneous counters we realize an aggregate metamaterial that
unambiguously detect a specific input `key' string and thus act as a sequential `lock'.
We note that state $\sigma=0$ for counter $bac$ is not unique to input $BAC$, but can
also be reached with input sequences such as $BBC$ and $CCC$ (Fig.~\ref{fig:4}c).
Hence, to uniquely recognize a string $BAC$, we combine the counting metamaterial $aaa|bbb|ccc$ with
the heterogeneous counter $bac$ (Fig.~\ref{fig:4}b,d). Out of all three-character strings, $BAC$ is the only one that yields the collective state $\{\sigma\}=\{0,1,2,0\}$ {(Fig.~\ref{fig:4}f)}. The experimental demonstration of the response of he $aaa|bbb|ccc|bac$ machine to input $BAC$ is shown in Fig.~\ref{fig:4}b,d, which correspond to a single experimental run where all four counters were actuated in parallel  (see Movie 6 in \cite{SI}). We note that our strategy can trivially be extended to longer sequences or larger alphabets.

While the design above cannot distinguish input $BAC$ from longer sequences such as $ABAC$,
we can detect such longer strings by extending the counter for the weakest signal: out of all possible input sequences, the metamaterial $aaaa|bbb|ccc|bac$ only reaches state $\{0,1,2,0\}$ for input $BAC$, thus allowing to uniquely filter and detect such a string. Finally we note that
designs featuring one heterogenous with multiple homogeneous counters are not  optimal. Unique detection of, e.g., three symbol sequences with less than four counters can be achieved; in addition, many
machines recognize multiple distinct input sequences (see Supplemental Material \cite{SI}).

\section{Outlook}

Our platform allows to realize metamaterials with predictable counting-like pathways and easily readable internal states under cyclical driving. These metamaterials act as a  sequential thresholding devices, and can be generalized to
detect more driving magnitudes and longer sequences. Moreover, similar sequential behavior can be realized in other designs, e.g., without slits.
In contrast to recent mechanical platforms that store mechanical bits \cite{pedro} and perform Boolean logic \cite{nature_mech_logic_gates,mechanical_computing_review}, our metamaterials perform
sequential computations, which are much more powerful than combinational logic.
Extending our update rules to more complex cases, including those where the new state depends on multiple neighbors, including in higher dimensions, opens routes to create
systems that are Turing-complete, such as 'rule 110' or
Conway's game of life \cite{automata,conway}.
Such 'cellular automata materials' would allow massively parallel
computations {\em in materia}.

\section*{Acknowledgements}
{We thank H. Bense, M. Caelen, C. Coulais, D. Holmes,
B. Dur\`a Faul\'i, D. Kraft, C. Meulblok and M. Munck for fruitful discussions and J. Mesman for technical support.}

\normalbaselines

\bibliographystyle{apsrev}

\cleardoublepage

\appendix

\begin{widetext}

{\Large Supplementary Information}
~\newline

\section{Sample Fabrication, Setup, Protocol, Design}
\begin{figure}[b]
	\begin{center}
		\includegraphics[width=.5\columnwidth]{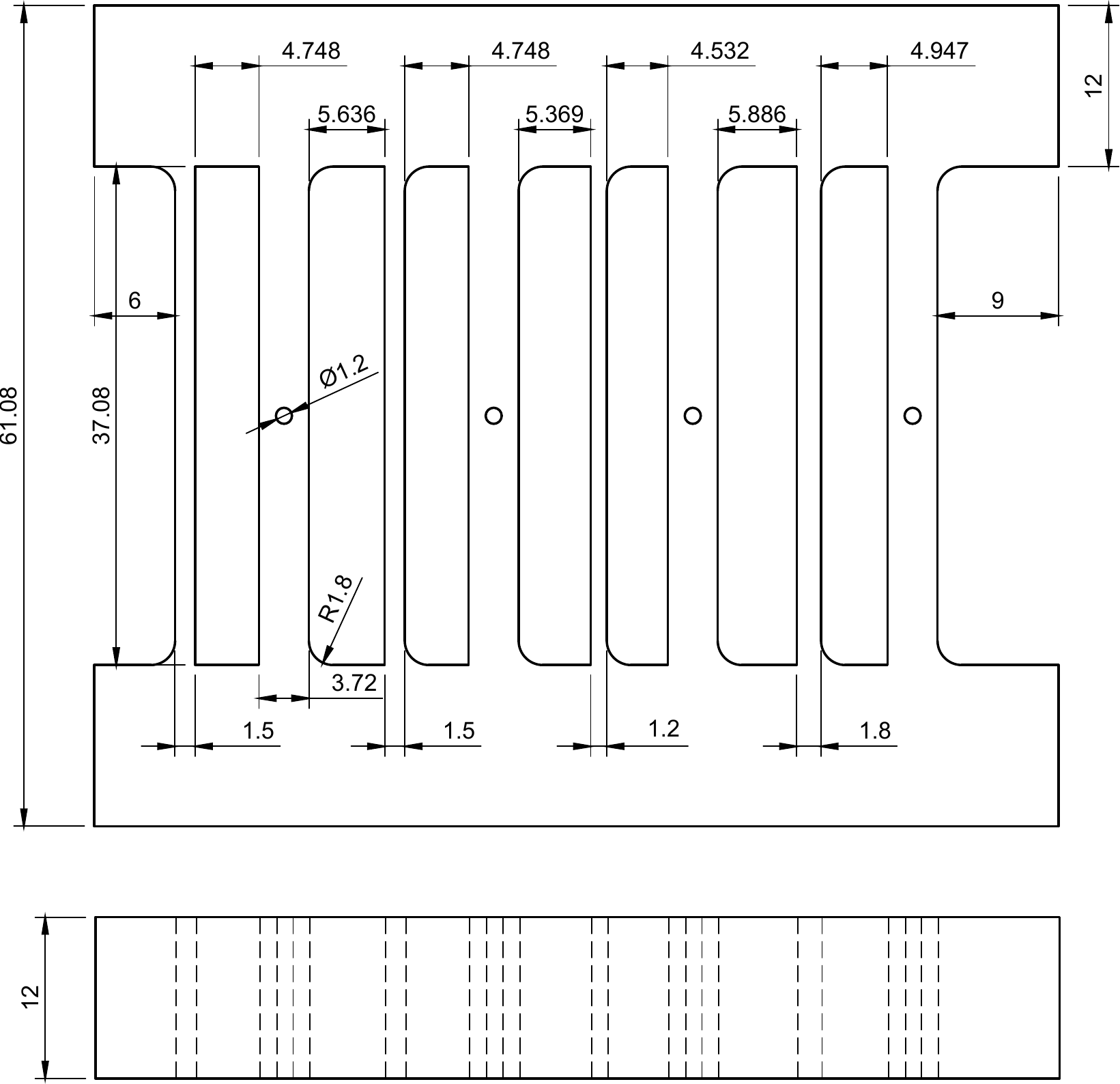}
	\end{center}
	\caption{Geometry of the $bac$ counter (distances in mm).}
	\label{fig:toscaledrawing}
\end{figure}

Samples are made by pouring degassed and refrigerated Zhermack Double Elite 32 VPS mixture into open face molds that are 3D printed on UltiMaker S3 and S5 printers. The samples are cured at room temperature. After curing, the samples are removed by breaking the molds. The slits in the s-beams are cut using a scalpel after which the samples are covered with talcum powder to reduce stiction. Before measurements the samples are allowed to rest for a week to allow their mechanical properties to settle.

The measurements are performed with a setup consisting of a homebuilt single-axis compression setup, where the samples are compressed between two plates, one static and one driven by a linear translation stage. The plates remain parallel within a slope of $<0.6\text{mm}/\text{m}$ and move at a rate of $<0.1\text{mm}/s$ with an accuracy of $\pm 0.01\text{mm}$.
A ccd camera captures images at a rate of $60 \text{Hz}$.

 \section{Robustness and Alternative Design}

\begin{figure}[h]
	\begin{center}
		\includegraphics[width=.8\columnwidth]{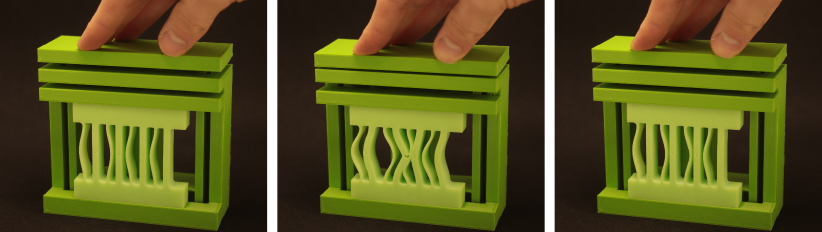}
	\end{center}
	\caption{A hand-operated flexure based device (dark green) containing a smaller
$n=4$ beam counter illustrates the  robustness of our design. }
	\label{fig:hh}
\end{figure}

\begin{figure}[b]
	\begin{center}
		\includegraphics{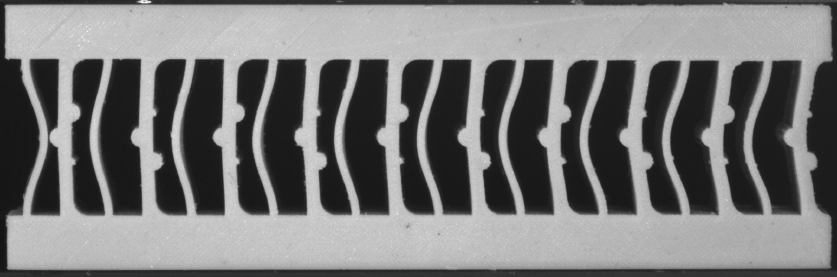}
	\end{center}
	\caption{Alternative design for beam counting using symmetry broken beams.}
	\label{fig:bumpy}
\end{figure}

To demonstrate that our design for homogeneous counters is robust, we produced scaled down versions that we place in a simple
hand-held device (see Fig.~\ref{fig:hh}). The device consists of a 3D printed flexure which guarantees a finite compression $\varepsilon_m$, and reasonably accurate parallel top and bottom plates. After setting the initial state to $\{1000\}$, repeated manual compression advances this counter towards its absorbing state $\{1111\}$  (Movie 2).

To demonstrate that beam counting can be realized in a wide variety of designs, we have explored an alternative design where the s-beams do not feature slits (see Fig.~\ref{fig:hh}). Here,
the symmetry breaking for counting is realized by pre-curvature of the m-beams, and
the addition of symmetry breaking bumps on the `s-beams'. The latter control the higher order buckling modes of the m-beams and guides their movement under cyclic compression (Movie 3).

\section{Design parameters}
All sample geometries have $L=37.1$~mm, $T=3.7$~mm, and a fixed radius of curvature of the symmetry breaking bases of $R=1.80$~mm (Fig.~\ref{fig:toscaledrawing}).
We fix $T$ and vary $t_i$, $d_i$ and $D_i$ to modify $\varepsilon^\dagger$ (Table.~\ref{table:dimensions}).
We measured $\varepsilon^{\dagger}$ by determining the rightward snapping of the relevant beams from the difference in subsequent images for each unit cell in our $aaa$, $bbb$, $ccc$ and $bac$ samples (Table.~\ref{table:epsdaggers}).
We note that there is scatter in the values of
$\varepsilon^{\dagger}$, and in particular that in the homogenous samples ($aaa$, $bbb$ and $ccc$), $\varepsilon^{\dagger}_1$ is lower than $\varepsilon^{\dagger}_2$ and $\varepsilon^{\dagger}_3$. We believe this to be due to the right-ward symmetry breaking of the left (0-th) m-beam in these samples. Not withstanding the scatter, our driving amplitudes bracket all the measured values for $\varepsilon^{\dagger}_a$, $\varepsilon^{\dagger}_b$ and $\varepsilon^{\dagger}_c$.

\begin{table}[h]
	\caption{Dimensions of homogeneous and heterogeneous samples}\label{table:dimensions}
	\begin{tabular}{l|c}
	Parameter                   & Dimension (mm) $\pm 0.05$ (mm) \\
	\hline
	m-beam thickness $t\left[a\right]$           & $1.2$          \\
	m-beam thickness$t\left[b\right]$			& $1.5$          \\
	m-beam thickness$t\left[c\right]$			& $1.8$          \\
	Beam length $L$                         & $37.1$          \\
	m-beam $\to$ s-beam distance $d\left[ a, \ldots \right]$ & $7.6$           \\
	m-beam $\to$ s-beam distance $d\left[ b, \ldots \right]$ & $7.9$           \\
	m-beam $\to$ s-beam distance $d\left[ c, \ldots \right]$ & $8.2$           \\
	s-beam $\to$ m-beam distance $D\left[\ldots, a\right]$ 	& $8.9$               \\
	s-beam $\to$ m-beam distance $D\left[\ldots, b\right]$	& $9.4$               \\
	s-beam $\to$ m-beam distance $D\left[\ldots, c\right]$	& $9.8$               \\
	Radius of slit-ending hole           & $0.6$                \\
	Radius of rounded corners $R$                  & $1.8$                \\
	\end{tabular}
\end{table}

\begin{table}[b!]
	\caption{Sample design and thresholds $\varepsilon^\dagger$.}
	\scriptsize
	\begin{tabular}{r|c|c|c|c|c|c|c}
	Sample
	 & $t_1$ $\pm 5\times10^{-4}$ (-)
	 & $t_2$ $\pm 5\times10^{-4}$ (-)
	 & $t_3$ $\pm 5\times10^{-4}$ (-)
	 & $t_4$ $\pm 5\times10^{-4}$ (-)
	 & $\varepsilon^\dagger_1$ {$\pm 0.5\times10^{-3}$} (-)
	 & $\varepsilon^\dagger_2$ {$\pm 0.5\times10^{-3}$} (-)
	 & $\varepsilon^\dagger_3$ {$\pm 0.5\times10^{-3}$} (-)\\
	\hline
	$aaa$  & $0.032$ & $0.032$ & $0.032$ & $0.032$ & $6.96\times10^{-2}$ & $7.58\times10^{-2}$ & $7.71\times10^{-2}$    \\
	$bbb$  & $0.040$ & $0.040$ & $0.040$ & $0.040$ & $8.17\times10^{-2}$ & $8.47\times10^{-2}$ & $8.76\times10^{-2}$   \\
	$ccc$  & $0.049$ & $0.049$ & $0.049$ & $0.049$ & $9.28\times10^{-2}$ & $9.44\times10^{-2}$ & $9.30\times10^{-2}$    \\
	$bac$  & $0.040$ & $0.040$ & $0.032$ & $0.049$ & $7.85\times10^{-2}$ & $7.58\times10^{-2}$ & $9.03\times10^{-2}$
	\end{tabular}
	\label{table:epsdaggers}
\end{table}

\section{Details of bit-copy operation}
Here we discuss in more detail the right copying of a '1' bit, as shown in Fig.~3 of the main paper --- see also movie 4.
During this process, the rightward buckled state of m$_1$ is copied to
s$_1$, and subsequently, the rightward buckled state of s$_1$ is copied to m$_2$. A detailed inspection of this dynamics further illustrates our design choices.
{\em(i)} First, upon increasing the compression from the initial state at $\varepsilon_m$ (main Fig.~3aI),
beam m$_1$ makes contact with the interaction beam s$_1$
before m$_2$: this is guaranteed by our choice of spacings $d<D$, and by $s_1$ buckling left due the rounded corners at its ends
(main Fig.~3aII). Increased compression beyond the critical strain $\varepsilon^\dagger$
results in the rightward snapping of s$_1$ --- this is why we take $d<D^*$ (see main Fig.~1c).
{\em(ii)} After s$_1$ has snapped, m$_2$ is sandwiched between s$_1$ and s$_2$, and takes on a complex shape (main Fig.~3aIII). Further compression does not lead to significant evolution. In this state, s$_1$ has a much larger deflection to the right than s$_2$ has to the left due to the presence of the slit, thus overcoming the difference between $D$ and $d$. This pushes
m$_2$ to the right, and when the strain is lowered, m$_2$ loses contact with s$_1$ and leans to the right (main Fig.~3aIV). We stress that the slits in the s-beams are crucial, as they enhance the rightward motion of s$_1$ --- without slits, the s-beams would push m$_2$ to the left as $d<D$ (see main Fig.~3b).
Upon further lowering of the strain to $\varepsilon_m$, the m-beam reaches a purely rightward buckled configuration and the system reaches state $\{11\}$ (main Fig.~3aV).
We have verified that all other initial conditions (\{00\}, \{01\}, \{11\}) remain invariant under the same cyclic driving --- as seen in main Fig.~1c. Hence, a judicious choice of geometry allows our unit cells to perform the irreversible rightwards advancing of 1-bits.

\section{Heterogeneous machines}

In the main text we presented a general strategy to
uniquely detect an input string of length $l_s=3$, featuring
$m=3$ characters, using $m$ homogeneous counters of length $l_s+1 (+2)$ and
one heterogeneous counter of length $l_s+1$.
This strategy can be extended to recognize strings of
arbitrary length, where $m$ homogeneous counters determine the multiplicity of each symbol, and a single heterogeneous counter uniquely reaches zero when its input matches its design.

However, often one can uniquely detect a given string with a aggregate metamaterial that features more than one heterogeneous counters.
To demonstrate this we performed an exhaustive search of the combinations of heterogeneous counters that uniquely detect input strings of length $l_s$ two and three, using $m=3$ characters, $\{A,B,C\}$. For every target string
we found a combination of $q$ counters, with $q\le l_s$ (Tables~\ref{tab:minimal2}-\ref{tab:minimal3}).
We note that some metamaterials can simplified even further, e.g., to detect $AA$
the metamaterial $aa|b$  suffices.

\begin{table}[!t]
	\caption{Examples of aggregate metamaterials consisting of two heterogeneous counters that reach a unique state for a given input string (key) of length two.
	}\label{tab:minimal2}
	\begin{tabular}{r|c|l}
		\toprule
		key & metamaterial &  unique state \\
		\midrule
		$AA$ &  $ba$ &  $[2]$ \\
		$AB$ &  $ac|ba$ &  $[1\; 1]$ \\
		$AC$ &  $ac|ba$ &  $[0\; 1]$ \\
		$BB$ &  $bb|ca$ &  $[0\; 2]$ \\
		$BC$ &  $ba|ca$ &  $[0\; 1]$ \\
		$CA$ &  $ab|ca$ &  $[1\; 0]$ \\
		$CB$ &  $ac|cb$ &  $[1\; 0]$ \\
		$CC$ &  $cc$ &  $[0]$ \\
		\bottomrule
	\end{tabular}
\end{table}

\begin{table*}[!t]
	\caption{Examples of aggregate metamaterials consisting of two heterogeneous counters that reach a unique state for a given input string (key) of length three.}\label{tab:minimal3}
	\begin{tabular}{ccc|ccc|ccc}
		\toprule
		key &  metamaterial      &     unique state &key  &  metamaterial       &     unique state&key  &  metamaterial       &     unique state \\
		\midrule
		$AAA$ &  $baa$ &  		$[3]$ 		&
		$AAB$ &  $aac|baa$ &  		$[1\; 2]$ 	&
		$AAC$ &  $aac|baa$ &  		$[0\; 2]$ 	\\
		$ABA$ &  $aab|aca|baa$ &  $[1\; 2\; 1]$ &
		$ABB$ &  $aab|aca|baa$ &  $[0\; 2\; 1]$ &
		$ABC$ &  $aca|baa$ &  		$[1\; 1]$ 	\\
		$ACA$ &  $aaa|aca|bba$ &  $[0\; 0\; 2]$ &
		$ACB$ &  $aaa|acb|bca$ &  $[0\; 0\; 2]$ &
		$ACC$ &  $aca|bca$ &  $[0\; 1]$ 		\\
		$BAA$ &  $aba|baa|caa$ &  $[2\; 0\; 3]$ &
		$BAB$ &  $aba|baa|caa$ &  $[1\; 0\; 3]$ &
		$BAC$ &  $aba|baa|caa$ &  $[1\; 0\; 2]$ \\
		$BBA$ &  $aab|bba|caa$ &  $[1\; 0\; 3]$ &
		$BBB$ &  $bbb|caa$ &  $[0\; 3]$ 		&
		$BBC$ &  $bba|caa$ &  $[0\; 2]$ 		\\
		$BCA$ &  $bca|cba$ &  $[0\; 2]$ 		&
		$BCB$ &  $bcb|cca$ &  $[0\; 2]$ 		&
		$BCC$ &  $bac|caa$ &  $[0\; 1]$ 		\\
		$CAA$ &  $aba|caa$ &  $[2\; 0]$ 		&
		$CAB$ &  $aba|cac$ &  $[1\; 1]$ 		&
		$CAC$ &  $aba|cac$ &  $[1\; 0]$ 		\\
		$CBA$ &  $aab|aca|cba$ &  $[1\; 2\; 0]$ &
		$CBB$ &  $aca|cbb$ &  $[2\; 0]$ 		&
		$CBC$ &  $aca|cba$ &  $[1\; 0]$ 		\\
		$CCA$ &  $aab|cca$ &  $[1\; 0]$ 		&
		$CCB$ &  $aac|ccb$ &  $[1\; 0]$ 		&
		$CCC$ &  $ccc$ &  $[0]$ 				\\
		\bottomrule
	\end{tabular}

\end{table*}

\begin{figure*}[t]
	\includegraphics[width=0.8\textwidth]{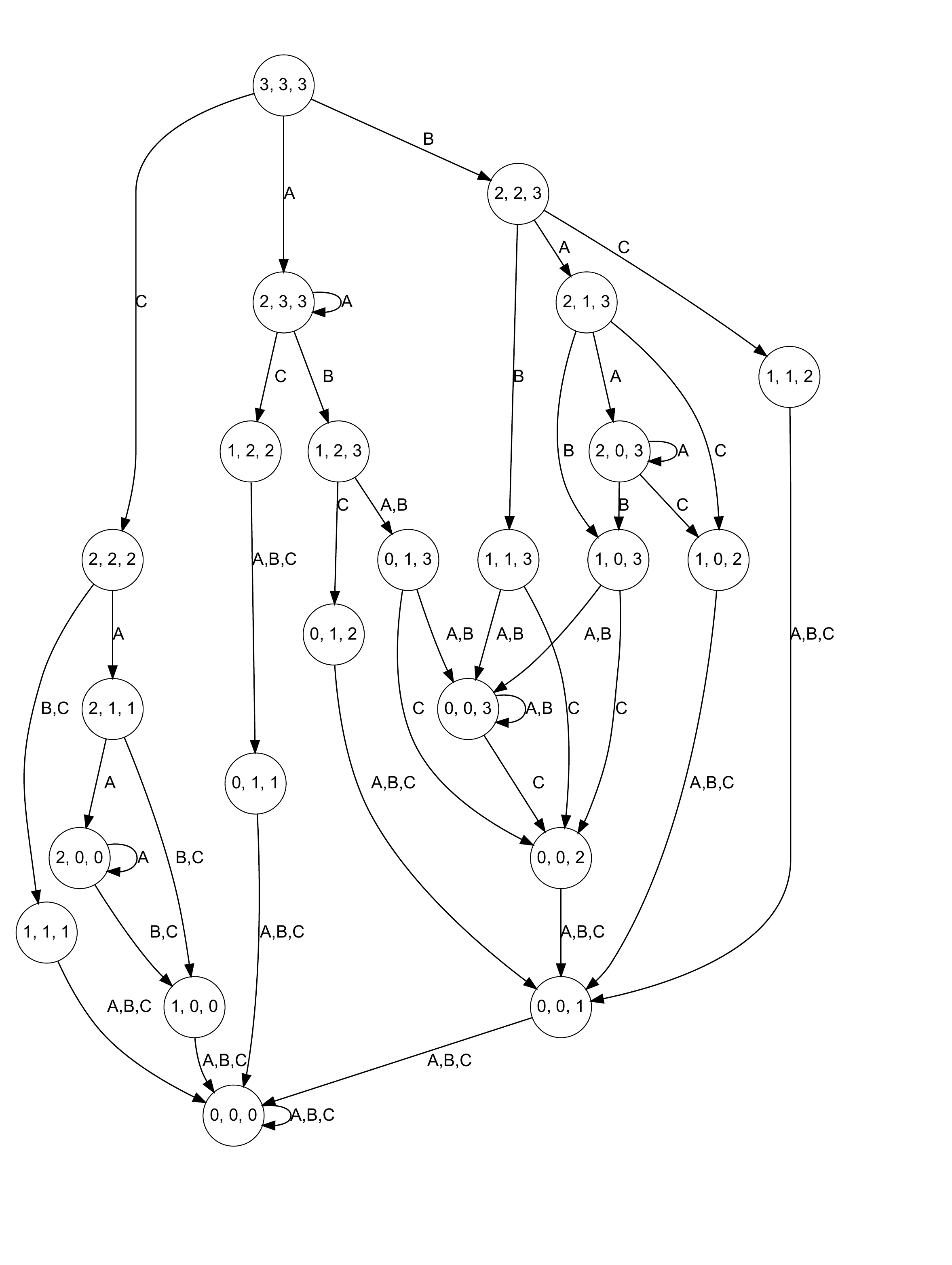}
	\caption{The full transition-graph for the aggregate metamaterial $aba|baa|caa$ (
labels are always to the right of the transition arrow).
	}
	\label{fig:aba|baa|caa}
\end{figure*}

{\bf Transitions in complex aggregate metamaterials}
To illustrate the complexity of metamaterials featuring multiple
heterogeneous counters, we show the transition graph for the aggregate metamaterial $aba|baa|caa$, starting out at state $\{3,3,3\}$ (Fig.~\ref{fig:aba|baa|caa}).
Clearly, the state that is reached encodes information on the driving sequence.
First, restricting ourselves to input strings of length three, there are eight states that encode a unique driving sequence:
$\{1, 0, 2\}, \{1, 2, 2\}, \{0, 0, 2\}, \{1, 2, 3\}, \{1, 0, 3\}, \{2, 3, 3\}, \{0, 1, 2\}, \{2, 0, 0\}$ and $ \{2, 0, 3\}$ --- 
as can be readily verified from Fig.~\ref{fig:aba|baa|caa}, each of these
is only reached in response to a  unique three-character input sequence. 
Second, taking input strings of arbitrary length into account, a different set of seven states is associated with a unique input sequence:
$\{3, 3, 3\}, \{2, 1, 1\}, \{2, 2, 3\}, \{1, 1, 2\}, \{1, 1, 3\}, \{2, 1, 3\}$ and $\{2, 2, 2\}$.

\clearpage
\section{Movie Captions}

\subsection*{Movie 1: Counting Metamaterial}
Evolution of the counting metamaterial shown in Fig.~2 of the main text. Notice that the movie starts at strain zero, and thus illustrates the effect of the symmetry breaking of
the m-beams in generating an initial state $\{10000000000\}$ at $\varepsilon_m$.

\subsection*{Movie 2: Hand-held Counter}
Evolution of a small four-cell metamaterial in a manually operated flexure-based compression device.

\subsection*{Movie 3: Other Starting Condition}
The 11-cel counter with initial condition shown in Fig.~2c of the main text.

\subsection*{Movie 4: Comparing with- and without slits}

Comparison of the evolution of two unit cells (original: left; design without slits: right) during a compression cycle. Note that we only cut the first s-beam. The difference images (A-B) illustrate the crucial effect of the slit in right copying the '1'-state of m-beams.

\subsection*{Movie 5: Alternative Design for Counting Metamaterial}

Example of a 10-unit cell metamaterial with an alternative design. After manually resetting its initial state to $\{1000000000\}$, we observe counting down from $\sigma=9$ to  $\sigma=0$ under cyclic compression.

\subsection*{Movie 6: $aaa|bbb|ccc|bac$ driven with $BAC$}
The aggregate metamaterial shown in Fig.~4 of the main text driven with the driving sequence $BAC$. The left three counters count the number of $A$, $B$ and $C$ signals, and the last counter only reaches the state $0$ (where all m-beams are to the right) for the specific permutation $BAC$.


\end{widetext}

\end{document}